\documentclass[aps,a4paper,twocolumn,showpacs]{revtex4}

\usepackage{graphicx}

\begin{document}

\newcommand{\be} {\begin{equation}}
\newcommand{\ee} {\end{equation}}
\newcommand{\ba} {\begin{eqnarray}}
\newcommand{\ea} {\end{eqnarray}}
\newcommand{\tr} {{\rm tr}}
\newcommand{\hata} {\hat{a}}

\title{Semiclassical Description of Wavepacket Revival} 

\author{Fabricio Toscano}
\email{toscano@if.ufrj.br}
\affiliation{ Instituto de F\'{\i}sica, 
              Universidade Federal do Rio de Janeiro, 
              Cx.~P. 68528, 21941-972 Rio de Janeiro, Brazil}

\author{Ra\'ul O. Vallejos}
\email{vallejos@cbpf.br}
\homepage{http://www.cbpf.br/~vallejos}
\affiliation{ Centro Brasileiro de Pesquisas F\'{\i}sicas (CBPF), 
              Rua Dr.~Xavier Sigaud 150, 
              22290-180 Rio de Janeiro, Brazil}

\author{Diego A. Wisniacki}
\email{wisniacki@df.uba.ar}
\affiliation{Departamento de F\'{\i}sica ``J. J. Giambiagi", 
             FCEN, Universidad de Buenos Aires, 
             1428 Buenos Aires, Argentina}

\date{\today}

\begin{abstract}
We test the ability of semiclassical theory to describe 
quantitatively the revival of quantum wavepackets
--a long time phenomena-- in the one dimensional quartic 
oscillator (a Kerr type Hamiltonian). 
Two semiclassical theories are considered: 
time-dependent WKB and Van Vleck propagation. 
We show that both approaches describe with impressive 
accuracy the autocorrelation function and wavefunction 
up to times longer than the revival time. 
Moreover, in the Van Vleck approach, we can show analytically 
that the range of agreement extends to arbitrary 
long times.
\end{abstract}


%
%

\maketitle

\section{Introduction}

From the early times of quantum theory there has
been a lot of interest in the possibility of 
describing quantum phenomena using approximate 
theories that take advantage of classical information 
\cite{maslov,brack}.
%
%
Not only because these {\em semiclassical} approaches 
provide in general a deeper insight into the system's 
behavior, due to the fact that classical quantities are more 
intuitive, but also in some cases the semiclassical 
computation of quantum quantities is made easier. 

The semiclassical propagation of wavefunctions
started with the seminal work of Van Vleck \cite{VanVleck}, 
where a semiclassical propagator was first introduced.
Gutzwiller \cite{Gutzwiller} showed that this propagator 
is the stationary-phase approximation of the Feymnan path 
integral, which results in a sum over classical paths.
At the same time Gutzwiller provided some corrections to 
Van Vleck's formula which are essential for long times.

The propagation of wavefunctions using the Van Vleck-Gutzwiller
propagator has fundamental problems that manifest themselves 
more sharply in systems with a classically chaotic dynamics.
The facts that, for long times, the integrations to be performed 
are usually highly oscillatory, and therefore not amenable to 
numerical 
computation \cite{brack}, and that the number of orbits 
becomes unmanageably large, has raised doubts about the long-time
semiclassical accuracy. 
These problems were controlled in Ref.~\cite{Tomsovic-Heller},
where a semiclassical method was applied to compute 
autocorrelation functions of Gaussian wavepackets.
This approach was shown to work both for chaotic 
\cite{Tomsovic-Heller} and regular systems 
\cite{Mallalieu_Stroud, Barnes} in 
long-time regimes but it has not been applied to calculate 
wavefunctions.

More recently, in Ref. \cite{toscano-raul-prl} it was shown that
Gaussian wavepackets can also be propagated by using the standard 
time dependent WKB theory (TDWKB) \cite{littlejohn92}, provided it is 
complemented with short-time methods.
In principle this scheme is suitable to compute wavefunctions 
in the long-time regime. 
Although initially conceived for chaotic systems, the scheme
also works for nonlinear integrable dynamics (see Sec.~IV below).

Among the semiclassical time-domain methods one should also 
highlight the ``initial value representation" (IVR)
of the propagator, especially suited for numerical 
implementations (see Ref.~\cite{miller2001} for a review).

In spite of the great advances in semiclassical theories in 
the last decades, there are still important open questions 
concerning the range of validity of semiclassical approximations 
and their accuracy in the description of subtle interference 
quantum effects such as, for example, quantum revivals.

The phenomenon of quantum revivals has been vastly studied in 
the literature (see Ref.~\cite{Robinett} for a review) and 
observed in many experiments, from atomic and molecular to 
optical systems \cite{Experiments}.  
The Wigner function of an initially well-localized wavepacket 
spreads for short times in a classical way, then it enters  
a delocalized quantum regime, and eventually it recombines 
itself to recover its original form. 
Such a revival occurs at a time, $T_{rev}$, which is long as
compared with classical timescales, like oscillation periods.
%
%
Perhaps more interestingly, in a wide class of circumstances, 
at times equal to a fraction of the revival time ($pT_{rev}/q$) 
the wavepacket relocalizes into a number of smaller copies of 
the initial packet, giving rise to ``fractional revivals". 
When the initial wavepacket can be associated to an essentially 
classical state, the ``fractional revival'' occurrence corresponds 
to a dynamical generation of Schr\"odinger's catlike states 
(a quantum superposition of macroscopically 
distinguishable states). 

One of the most interesting systems exhibiting both full and 
fractional revivals is the quartic oscillator, whose Hamiltonian 
is obtained by the squaring the harmonic oscillator Hamiltonian.
In essence this is the Hamiltonian of the Kerr model,  
describing a single mode of the quantized radiation field in
a nonlinear medium, and extensively studied in quantum optics. 
The formation of revivals and fractional revivals in this
system was analyzed, 
for example, in \cite{revival-kerr-medium}.
Recently the quartic oscillator has experienced a renewed 
interest for its connections with quantum information processing 
in continuous variable (CV) systems.
Indeed, Hamiltonians of the Kerr type are the simplest nonlinear 
ones acting on a single mode ({\it i.e.} one-mode quantum logic gate) 
needed to define universal quantum computation within 
the subclass of unitary transformations generated by 
Hamiltonians that are polynomial functions of the CV operators 
\cite{braunstein2005}. 

In addition to the already mentioned studies of revivals in
the Coulomb problem \cite{Mallalieu_Stroud, Barnes}, we must
also mention the papers by Wang and Heller \cite{Wang-Heller}
and by Novaes \cite{Novaes}.
The first authors considered the revival of a wavepacket in 
the Morse potential. 
Using a convenient numerical implementation of the 
Van Vleck-Gutzwiller propagator they succeeded in reproducing 
satisfactorily the first revival of the wavefunction. 
%
%
Novaes studied the semiclassical propagation of a wavepacket 
in the quartic oscillator, starting from the semiclassical
coherent-state representation of the propagator. 
Even though he also obtained an excellent agreement, the 
increasing difficulty in determining the required complex 
trajectories as time grows limited the application of the 
method to short times (a few classical periods) 
\cite{Novaes}. 

The present paper is devoted to show that  ``elementary"  
semiclassical theories can be successfully applied to describe 
the revival phenomena in the quartic oscillator. 
Our study focus both on the autocorrelation function and 
on the wavefunction.
The two elementary semiclassical theories examined are two:
Van Vleck propagation (Sec.~III) and 
time-dependent WKB (Sec.~IV). 
In the first case, calculations are analytical; in the second,
numerical.
In both cases we find an excellent agreement between semiclassical
theory and exact propagation even at very long times 
(e.g., multiples of the revival time). 
In the particular case of the Van Vleck autocorrelation function
we prove analytically that it agrees with the exact one up to
arbitrary long times, the error being semiclassically small and
independent of time.
As a byproduct of our study, we show that TDWKB also works
efficiently in integrable nonlinear systems 
\cite{toscano-raul-prl}.


Section~II contains a description of the main aspects of the 
model.
We present our main conclusions in Sec.~V.

\section{The model}

Consider a one degree of freedom harmonic oscillator:
\be
H=\frac{\hat{p}^2}{2m}+ \frac{1}{2} m \omega^2 {\hat q}^2 \,.
\ee
Throughout the paper we set $m=1$ and $\omega=1$ (if desired,
these constants can be recovered at any moment by dimensional
considerations).
We shall be interested in the dynamics of coherent states of 
this harmonic oscillator, i.e., eigenstates of the annihilation 
operator \cite{cohen}
\be
\hat{a}=\frac{1}{\sqrt{2 \hbar}}(\hat{q}+i\hat{p}) \,,
\ee
with  $[\hat{a},\hat{a}^{\dagger}]=1$ ($\hat{a}^{\dagger}$ the 
creation operator).
Our model to study the revivals and fractional revivals 
is the quartic Hamiltonian given by,
\be
\label{1st-quantum-Hamiltonian}
\hat{H}= \gamma \hbar^2 \left (\hata^\dagger \hata + 
         \frac{1}{2} \right)^2 
         \equiv \gamma \hbar^2 \left (\hat{n} + 
         \frac{1}{2} \right)^2 \,,
\ee
where $\hat{n}=\hat{a}^{\dagger}\hat{a}$ is the number operator.
This nonlinear Hamiltonian is of the Kerr type \cite{milburn1995}.
Recall that the effective Hamiltonian that describes the dynamics 
of a single light mode inside a high finesse optical cavity
containing a Kerr medium is: 
\be
\hat{H}= \omega_1 \hat{n}^2 -\omega'_1 \hat{n}   
\ee
($\omega_1$ and $\omega'_1$ are real frequencies), and we 
recover the
optical context  in our formalism considering  $\hbar=1$.

%
%

\begin{figure*}[htp]
\setlength{\unitlength}{1cm}
\begin{center}
\scalebox{0.95}[0.95]{%
\includegraphics*[width=8.0cm,angle=-90]{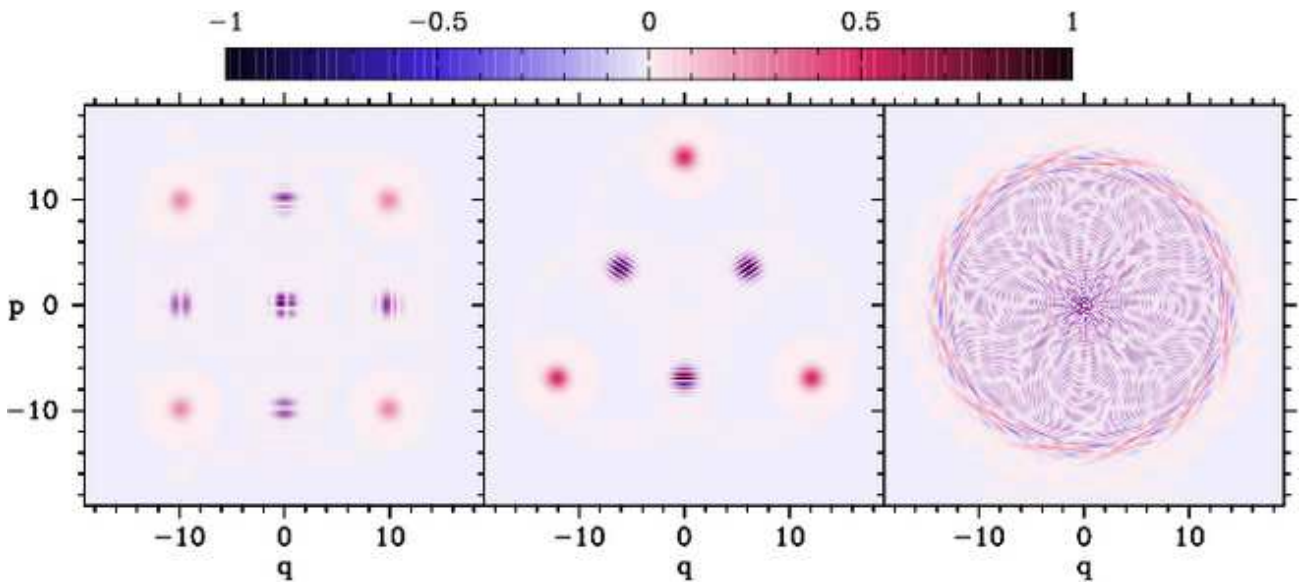}}
\end{center}
\caption{(Color online) 
Snapshots of the evolution of an initially Gaussian wavepacket 
(Wigner functions). 
From left to right, times are $t=T_2/4, \, T_2/3, \, T_2/2.38567$,
with $T_2$ the revival time.
The initial coherent state is centered at $(q_0,p_0)=(0,14)$.}
\label{fig1}
\end{figure*}

The quantum evolution of a coherent state, 
$| \alpha_0 \rangle $, with the Hamiltonian
(\ref{1st-quantum-Hamiltonian}) yields
\be
|\psi\rangle=e^{-i\hat{H}t/\hbar} 
              | \alpha_0 \rangle= e^{-|\alpha_0|^2/2} \sum_{n=0}^\infty 
        \frac{|\alpha_0|^{2n}}{\sqrt{n!}}
        e^{-i \gamma \hbar (n+1/2)^2 t } |n\rangle\,,
\label{psiex}
\ee
where $|n\rangle$ is a number state, {\it i.e.,} 
$\hat{n}|n\rangle=n|n\rangle$.
At multiples of the revival time
\be
\label{revival-time}
T_2= \frac{\pi}{\gamma \hbar}
\ee
the dynamics reconstructs the initial coherent state.
For times equal to $(p/q)T_2$ ($p/q$ an irreducible fraction) 
the evolved state consists of a 
superposition of $q$ 
coherent states lying on a circle of radius $|\alpha_0|$ 
(fractional revivals) \cite{revival-kerr-medium}.
In Fig.~\ref{fig1} we show the Wigner function \cite{schleich} 
of the evolved state (\ref{psiex}) at several selected times. 
Schr\"odinger catlike states can be clearly seen at
the fractional revival times.
%
%

We shall focus on the autocorrelation function, 
\ba
C_1(t)&=& \langle \alpha_0 | 
            e^{-i\hat{H}t/\hbar} 
              | \alpha_0 \rangle \nonumber\\
           &=&   e^{-|\alpha_0|^2} \sum_{n=0}^\infty 
        \frac{|\alpha_0|^{2n}}{n!}
        e^{-i \gamma \hbar (n+1/2)^2 t } \,,
\label{C1ex}
\ea
in the semiclassical regime, defined by the condition
$|\alpha_0|^2 \gg 1$,
which is the appropriate semiclassical limit in the case 
of an optical context.
It is important to note that this function shows the 
periodicity property
\be
C_1(t+T_2)= e^{-i \pi/4} \, C_1(t) \, ,
\ee
the period being the revival time (\ref{revival-time}).

Note that the Weyl-Wigner representation 
\cite{schleich,report-alfredo} of the Hamiltonian
(\ref{1st-quantum-Hamiltonian}) is: 
$\gamma[(q^2+p^2)/2]^2-\hbar^2/4$.
Thus, the classical Hamiltonian corresponds to
\be
H(q,p)= \gamma I^2 \,,
\label{ham}
\ee
with the action variable
%
%
%
\be
I(q,p)= 
\frac{1}{2\pi} \oint \;p \;dq = 
\frac{1}{2   } \left( q^2 + p^2 \right) \,,
\ee
a constant of motion.
%

\section{Van Vleck approach}

The semiclassical calculations in this section are based 
on the Van Vleck-Gutzwiller approximation 
\cite{VanVleck,Gutzwiller} to the propagator:
\be
K(q'',q',t) 
\approx 
\frac{e^{-i \pi/4}}{\sqrt{2 \pi \hbar}} 
\sum_k A_k
e^{iS_k(q'',q',t)/\hbar-i\mu_k\pi/2} \,.
\label{vv}
\ee
The sum runs over classical trajectories connecting 
$q'$ to $q''$ in time $t$. 
Each trajectory contributes with an amplitude $A_k$ and a 
phase. 
The phase is made up from the Lagrangian action $S_k$,
\be
S_k(q'',q',t)=\int_0^t (p\dot{q}-H) dt \, ,
\ee
and the Maslov index $\mu_k$, which (in the present case)
coincides with the number of turning points (where $\dot{q}=0$)
encountered by the trajectory \cite{littlejohn92}.
The amplitude is given by
\be
A_k=\frac{1}{\sqrt{\left|\partial q''/\partial p' \right|_k}} \,,
\ee
with 
\be
q''=q''(q',p',t) \,.
\ee
Equation~(\ref{vv}) can be derived, for instance, by 
using time-dependent WKB theory to propagate a position
eigenstate \cite{littlejohn92}.

\subsection{Autocorrelation function}

Consider the autocorrelation function
\be
C_2(t)= \int_{-\infty}^{\infty} \int_{-\infty}^{\infty} 
         dq'' dq' 
         \, \psi_0^\ast(q'') 
         \, \psi_0(q') 
         \, K(q'',q',t) \,,
\label{C2t}
\ee
where $\psi_0(q)$ is the wavefunction of the initial state. 
By substituting the exact propagator $K(q'',q',t)$ by 
Van Vleck's we obtain a semiclassical correlation function.

The simplicity of the system we are considering permits
the analytical determination of the required trajectories; 
the properties of the trajectories can thus be calculated to 
the desired precision.
The classical equations of motion are obtained from
the Hamiltonian (\ref{ham}):
\ba
q(t) & = & \sqrt{2I} \sin[ \beta + \omega(I) t ] \, , \\
p(t) & = & \sqrt{2I} \cos[ \beta + \omega(I) t ] \, ,
\ea
with
\be
\omega(I)= \frac{\partial H}{\partial I} = 2 \gamma I \, .
\ee
Both the action $I$ --a constant of motion-- and $\beta$ are 
determined by the initial conditions.
The trajectories contributing to the Van Vleck propagator are 
the solutions of the following boundary problem:
\ba
\label{qq1}
q'' & = & \sqrt{2I} \sin[ \beta + \omega(I) t ] \, , \\
q'  & = & \sqrt{2I} \sin(\beta) \, ,
\label{qq2}
\ea
for a given time $t$.

Among all trajectories that satisfy the equations above,
only a small subset will be relevant to the autocorrelation 
function. 
The reason is that the initial wavepacket is localized on a 
phase-space region of radius ${\cal O}(\sqrt{\hbar})$,
meaning that the important trajectories are only those having 
endpoints 
$(q'',p'')$ and $(q',p')$ 
in a region of radius 
${\cal O}(\sqrt{\hbar})$ 
around the center of the wavepacket.

We shall consider, for convenience (but without loss of
generality), an initial wavepacket corresponding to a coherent
state $|\alpha_0\rangle$ located on the $p$-axis, i.e., $\alpha_0=ip_0/\sqrt{2 \hbar}$, 
whose wavefunction is
\be
\psi_0(q)=( \pi \hbar)^{-1/4} 
                  e^{-q^2 / 2 \hbar}
                  e^{i p_0 q/\hbar} \,.
\ee
In this case, we can assume that both $q''$ and $q'$ are
small [${\cal O}(\sqrt{\hbar})$] and resort to Taylor
expansions around $(q_0=0,p_0)$.
We can get a good idea of the structure of the set of 
trajectories that we need by analyzing the particular case $q''=q'=0$. 
From Eqs.~(\ref{qq1},\ref{qq2}) we obtain a set of periodic
trajectories, labeled by the winding number $k$, the 
corresponding actions being
\be
I_k= \frac{p_k^2}{2}=
\frac{k \pi}{\gamma t} \, .
\label{Ikpk}
\ee
The relevant values of $k$ are those satisfying 
\be
I_k \approx I_0  \equiv \frac{p_0^2}{2} \, ,
\ee
i.e.,
\be
k \approx k_0 \equiv \frac{p_0^2 \gamma t}{2 \pi}  \, .
\ee
For simplicity we ignore $k=0$ trajectories, which would
require a special treatment.
This means that our calculation is not valid for times 
of the order or smaller than $T_1$,
where $T_1$ is the period of the classical motion of
the centroid of the wavepacket:
\be
T_1 = \frac{2 \pi}{\omega_0} \equiv 
      \frac{2 \pi}{2 \gamma I_0} =
      \frac{  \pi}{  \gamma I_0}  \, .
\ee
(This is not a problem as we are interested in long times.)
The number of trajectories that contribute at $t \approx k T_1$
is of the order of $\sqrt{k}$ (see Sect.~\ref{secIVC}).
The properties of these periodic trajectories are 
easily calculated:
\ba
\mu_k^{(0)}  & = & 2k                              \, , \\
  S_k^{(0)}  & = & 2k \pi  I_k - H(I_k) t = 
             \frac{\pi^2 k^2}{\gamma t}            \, , \\
  A_k^{(0)}  & = & \frac{1}{\sqrt{4I_k \gamma t}}  \, .
\ea

In the general case, when $q''$ and/or $q'$ are not strictly 
zero, even if the corresponding trajectories are not periodic 
any more, they can be put into one-to-one correspondence with
the periodic ones.
Accordingly, the new actions and amplitudes can be calculated
as Taylor expansions in $q''$ and $q'$ around the periodic
solutions. 
(The Maslov indices do not change, as they just count twice 
the number of turns.) 

We shall approximate the actions $S_k$ to second order in 
$(q'',q')$ and the amplitudes $A_k$ will be kept at zero-th order
\cite{Tomsovic-Heller,Barnes}.
So, we need to calculate first and second derivatives of $S_k$ 
with respect to $q''$ and $q'$ and evaluate them at 
$(q''=0,q'=0,t)$. 
The results are
\ba
\frac{\partial S_k}{\partial q'' } & = & 
-\frac{\partial S_k}{\partial q' } = p_k \, , \\
\frac{\partial S_k^2}{\partial q''^{\, 2} } & = & 
\frac{\partial S_k^2}{\partial q'^{\, 2} } =
-\frac{\partial S_k^2}{\partial q' \partial q'' }=
\frac{1}{4 I_k \gamma t} \,. 
\ea
Thus, we arrive at
\ba 
S_k & \approx & \frac{\pi^2 k^2}{\gamma t}  + 
                \sqrt{\frac{ 2\pi k}{\gamma t}}\,(q''-q') + 
                \frac{(q''-q')^2 }{8k\pi}  \, , \\
A_k & \approx &
              A_k^{(0)} = \frac{1}{\sqrt{4I_k \gamma t}} \, . 
\ea
The final steps in the calculation of the semiclassical 
correlation function are: 
  (i) substitute the expressions above into 
      Van Vleck propagator (\ref{vv}), 
 (ii) insert the resulting propagator into the definition
      of correlation function (\ref{C2t}),
(iii) calculate the Gaussian integrals.
In this way we obtain:
\be
C_2(t)  \approx 
\frac{ e^{-i\pi/4} }
     { \sqrt{2 \pi } }
\sum_{k=1}^\infty 
       e^{i \pi^2 k^2/ \gamma \hbar t} e^{-ik \pi} 
\,
\frac{\,  e^{-\left( p_k-p_0 \right)^2/a_k \hbar}}
     {\sqrt{ k \, a_k}} \, . 
\label{corrVV}
\ee
with
\be
a_k=1-\frac{i}{2 k \pi}  \, ,
\ee
and $p_k=\sqrt{2k\pi/\gamma t\/}$.

Equation~(\ref{corrVV}) gives the semiclassical correlation 
function obtained from Van Vleck's propagator and perturbation 
analysis around periodic orbits; it is a well behaved sum that
can, in principle, be calculated for any $t$ and compared with
the exact result.
The domain of validity is $t \gtrsim T_1$.
Now we proceed to numerical comparisons, postponing analytical
considerations to Sect.~\ref{secIVC}.

\begin{figure}[htp]
\hspace{0.0cm}
\includegraphics[angle=0.0, width=8cm]{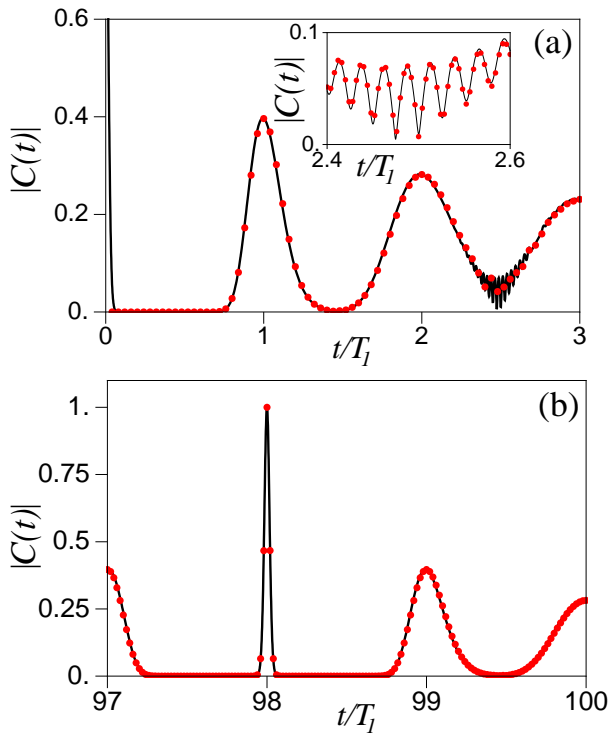}
\caption{%
Absolute value of the autocorrelation function vs. time, in 
units of the classical period $T_1$. Line: exact; dots: 
semiclassical approximation. 
We set $m,\omega,\hbar,\gamma=1$, $p_0=14$.
(a) Short times. Inset: blowup of the region where
    the first interference effects appear. 
(b) Times around the second revival ($t \approx 2T_2$).}
\label{figCorr}
\end{figure}

In Fig.~\ref{figCorr}, we display semiclassical and exact
correlation functions in two time windows, one for short
times [panel (a)], the other for times around the second 
revival [panel (b)]. 
Except for the initial correlation peak ($t \approx 0$),
where the semiclassical approximation was expected to fail,
the agreement is excellent. 
Remarkably, the peak that is lost at $t \approx 0$, 
it perfectly resurges at $t \approx 2T_2$
(and at other revival times --graphics not shown).
%
%
%

\subsection{The Van Vleck wavefunction}
\label{secVV}

There is a straightforward test for the wavefunctions
generated by the Van Vleck approach:
If, instead of doing both the integrals defining the
correlation function in (\ref{C2t}), we calculate only 
the first one, over $q'$, we obtain a Van Vleck 
wavefunction $\psi(q'')$.
First of all, this semiclassical wavefunction can be 
tested only in a small interval around $q''=0$ 
(because of the approximations we did in calculating 
the Van Vleck propagator). 
Moreover, this calculation omits the contributions
arising from trajectories that arrive at $q''$ in
time $t$ {\em with negative momentum}.
So, we expect the approximation to work well only
at times where the (Wigner function of the) evolved 
state does not have negative momentum components 
around $q''=0$.
Such a situation arises, for instance, at revival
times, when the initial wavepackets is exactly
reconstructed.

Another cases that can be described in terms of 
periodic orbits are fractional revivals like that
at $t=T_2/3$ (see middle panel of Fig.~\ref{fig1}), 
when the wavepacket is a superposition of three coherent 
states with phases 
$\arg(\alpha)=\pi/2,2\pi/3,4\pi/3$. 
In this case one has:
\be
|\psi(t)\rangle = e^{-i \pi/12} \, c_0 \, 
|\alpha_0 \rangle + \ldots \,,
\label{eqfrac}
\ee
where 
\be
c_0 = \frac{2+e^{-2\pi i/3}}{3} \, ,
\ee
and the ellipsis stand for the other two coherent 
states.

\begin{figure}[htp]
\hspace{0.0cm}
\includegraphics[angle=0.0, width=8cm]{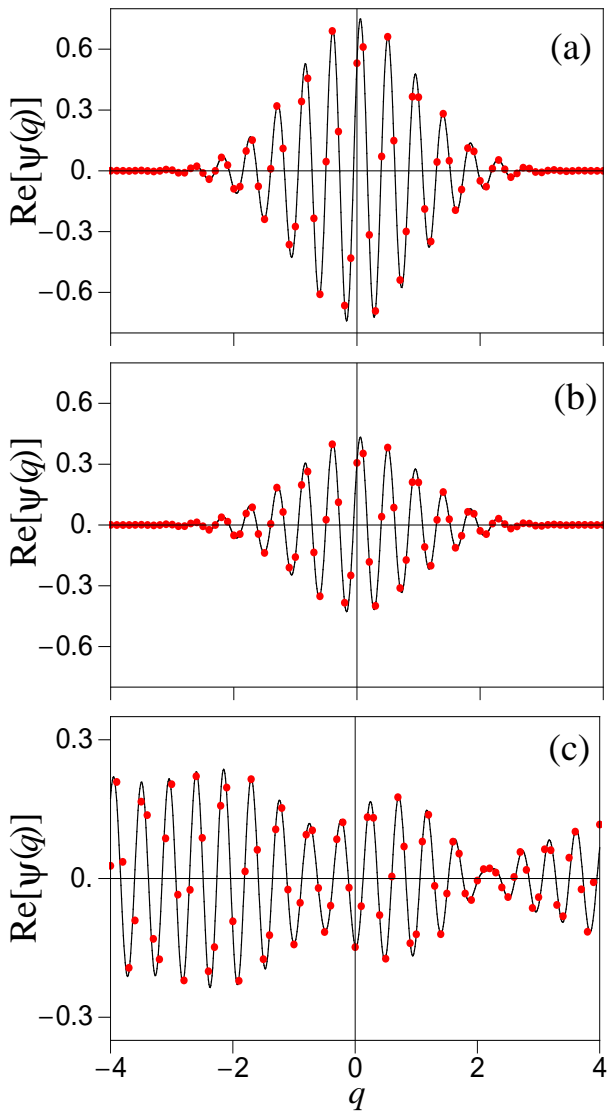}
\caption{%
Real part of the wavefunction $\psi(q)$ vs. $q$ at some
selected times. Line: exact; dots: Van Vleck. 
Similar agreement is observed for the imaginary part.
We used $m=\omega=\hbar=\gamma=1$, $p_0=14$.
(a) At the first revival time. 
(b) At the fractional revival time $t=T_2/3$.
(c) At $t=T_2/2.38567$, when the wavefunction is completely
    delocalized.}
%
\label{figWave}
\end{figure}

Figure~\ref{figWave} [(a) and (b)] shows that indeed 
the Van Vleck scheme which only uses periodic orbits
reproduces the exact wavefunctions almost perfectly at 
times $t=T_2$ and $t=T_2/3$.
%
%
%
%

In order to calculate the wavefunction at an arbitrary 
time (but not too short),
we must also take into account the contributions of the
family of trajectories that, starting at $q'$, arrive 
at $q''$ in time $t$ with negative momentum.
As done before, we Taylor-expand actions and amplitudes
around the main family $(q''=0,q'=0)$.
These are ``half-periodic" trajectories, i.e., their 
frequencies satisfy $\omega_k t= 2 \pi (k-1/2)$, for 
$k \ge 1$. 
We skip the details and just show the final results:

\ba 
  S_k & \approx &  \frac{\pi^2 k^2}{\gamma t}  - 
                \sqrt{\frac{ 2\pi k}{\gamma t}}\,(q''+q') + 
                \frac{(q''+q')^2 }{8k\pi}  \, , \\
  A_k & \approx & \frac{1}{\sqrt{4I_k \gamma t}}   \, , \\
\mu_k &    =    & 2k-1  \, .
\ea 
Adding the contributions of the family above to that
of the periodic trajectories we obtain a semiclassical 
wavefunction valid for arbitrary times.
Figure~\ref{figWave} (bottom panel) compares
semiclassical and exact wavefunctions at a time when 
the state is completely delocalized in phase space
(rightmost panel in Fig.~\ref{fig1}).
Again, the semiclassical approximation performs remarkably 
well for small enough values of $q$.
Not unexpectedly, for $|q| \approx 3$ some small deviations 
start to show up, and keep growing with increasing $|q|$.

\subsection{Analytical comparisons}
\label{secIVC}

The striking accuracy of the semiclassical approximations
verified in the previous sections, together with the 
relative simplicity of the expressions involved,
suggests that it should be possible to give an analytical
proof of the approximate quantum-semiclassical equivalence.
In the following we show analytically that the error committed
in the semiclassical autocorrelation function is really very
small and, what is especially important, independent of time. 
(A similar analysis could in principle be carried out 
for wavefunctions but will not be attempted here.)

When comparing semiclassical (\ref{corrVV}) and exact (\ref{C1ex}) 
correlation functions we immediately see a fundamental
difference: time appears in a different way in both expressions. 
While the quantum expression is a Fourier series, i.e., a sum of
plane waves (in $t$), the semiclassical correlation is a sum of wavepackets, or wavetrains.  
The tool that switches between wavetrains and plane 
waves is Poisson transformation \cite{schleich}.
But before applying Poisson transformation to, say, the quantum
correlation function let us introduce a small simplification.

\subsubsection{Quantum correlation function}

For large $|\alpha_0|^2$ one can approximate the Poisson 
distribution by a Gaussian \cite{schleich}:
\be
    e^{-|\alpha_0|^2} \frac{|\alpha_0|^{2n}}{n!} \approx
         \frac{1}{\sqrt{2\pi\nu}} 
         e^{-(n-\nu+1/2)^2/2\nu} \,,
\ee
where  $\nu =|\alpha_0|^2$. 
With this approximation the exact autocorrelation can be 
written in terms of the Jacobi theta function \cite{ww}
\be
\vartheta_3(z|\tau)= \sum_{n=-\infty}^\infty  
            e^{\pi i n^2 \tau + 2 i n z} \,.
\ee
In fact, after the substitution Poisson-to-Gaussian in 
Eq.~(\ref{C1ex}), we extend the lower limit of the summation 
to $-\infty$, obtaining
\be
C_1(t) \approx 
         \frac{1}{\sqrt{2\pi\nu}} \,
         e^{-(\nu-1/2)^2/2\nu}     \,
         e^{-i \gamma \hbar t/4 } \,
        \vartheta_3(z_1 | \tau_1) \,,
\label{C1ap}
\ee
where
\ba
z_1    & = & \frac{1}{2  i} 
            \left( 1 -\frac{1}{2\nu}-i \gamma \hbar t  
            \right) \, , \\
\tau_1 & = & \frac{1}{ \pi i} 
            \left( -\frac{1}{2\nu}-i \gamma \hbar t  
            \right) \, .
\ea

Expressing $C_1(t)$ in terms of 
$\vartheta_3$ brings
in several benefits. 
First, formulas become more compact and can be calculated 
quickly and efficiently using standard softwares, e.g., 
Mathematica \cite{wolframtheta}, where $\vartheta_3$ is a 
built-in function. 
Second, Poisson transforming the theta function is equivalent 
to using the {\em functional equation}
\be
\vartheta_3(z|\tau)=
(-i \tau)^{-1/2} 
e^{z^2/\pi i \tau}
\vartheta_3 \left( \left. \frac{z}{\tau}  \right| 
                         -\frac{1}{\tau} \right) \, ,
\ee
where $(-i \tau)^{-1/2}$ is to be interpreted by the 
convention $|\arg(-i \tau)|<\pi/2$ \cite{ww}.
(This equation has already been used by Wang and Heller
in their semiclassical study of the square well
\cite{Wang-Heller}.)

Now we use the functional equation 
but restricting 
ourselves to times longer than the period $T_1$, i.e.,  
$\gamma \hbar t \gg \nu$. So,
\ba
 \frac{z_1}{\tau_1} 
                & \equiv & z_1^\prime 
                    = \frac{\pi}{2} + 
                      \frac{i \pi}{2 \gamma \hbar t} + 
                      \ldots \, , \\ 
-\frac{1}{\tau_1} 
                & \equiv & \tau_1^\prime 
                    = \frac{\pi}{\gamma \hbar t } +
                      \frac{i \pi}{2 \nu \gamma^2 \hbar^2 t^2 } +
                      \ldots \, , \\
\frac{z_1^2}{\pi i \tau_1} 
                & = & \frac{i \gamma \hbar t }{4} 
                    - \frac{1}{2} + \frac{1}{8 \nu}+
                       \ldots \, , \\
(-i \tau_1)^{-1/2} 
                & = & e^{-i \pi/4}
                      \sqrt{\frac{\pi}{ \gamma \hbar t }} + 
                       \ldots \, . 
\ea
In this way we arrive at an approximate expression for the 
quantum correlation function which has a semiclassical 
appearance:
\be
C_1(t) \approx 
         \frac{e^{-\nu/2}}{\sqrt{2 \nu \gamma \hbar t}} \,
         e^{-i \pi/4 } \,
        \vartheta_3(z_1^\prime | \tau_1^\prime) \,.
\label{C1ap2}
\ee
Before comparing this expression with the semiclassical
one we shall do some manipulations of the semiclassical
formula.

\subsubsection{Van Vleck correlation function}

The semiclassical correlation function is written as a
sum over periodic orbits. 
Each orbit is weighted by the exponential function:
\be
W(k) \equiv 
\frac{\, \, e^{-\left( p_k-p_0 \right)^2/a_k \hbar}}
     {\sqrt{k \, a_k}}  \, .
\label{Wk}
\ee
Recall that this function depends on time through 
$p_k$ (\ref{Ikpk}). 
For long times (large $k$) we can set $a_k \approx 1$.
Thus $W(k)$ becomes a real function, which can be well
approximated by a Gaussian.

\begin{figure}[htp]
\hspace{0.0cm}
\includegraphics[angle=0.0, width=8cm]{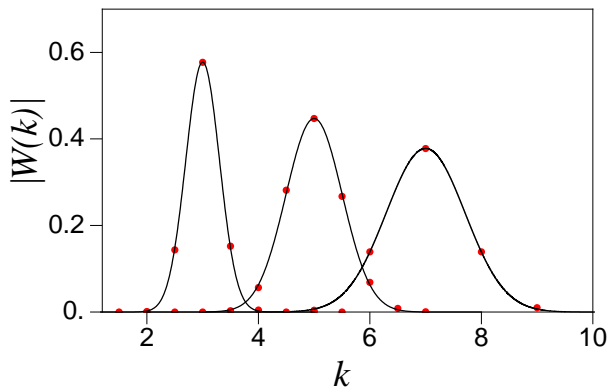}
\caption{%
Orbit weight function for times $t/T_1=3,5,7$
(from left to right, respectively).
Lines: Gaussian approximation, with $k$ a continuous 
variable.
Dots: absolute value of exact weight.
Parameters are $m,\omega,\gamma,\hbar=1$, $p_0=14$.}
\label{figW}
\end{figure}

We show in Fig.~\ref{figW} some plots of the exact 
$W(k)$ [Eq.~(\ref{Wk})] 
for typical values of the parameters together with
the simplest Gaussian approximation obtained by setting 
$k=k_0$, $a_k=1$ in the denominator, and linear expanding
$p_k$ around $k=k_0$, i.e.,
\be
W(k) \approx
\frac{\, \, e^{-\pi\left( k-k_0 \right)^2/ 
                   2 k_0 \gamma \hbar t}}
     {\sqrt{k_0}}  \, .
\label{appW}
\ee
%
%
%
%
%
It is verified that the Gaussian approximation is already 
very good for the small values of $k$ considered.

The remarkable fact is that if we substitute the approximate
weight function (\ref{appW}) into the semiclassical
correlation function (\ref{corrVV}) we obtain the 
approximate quantum correlation function (\ref{C1ap2}).
This concludes the analytical comparison of quantum and
semiclassical correlation functions: We have shown that 
for long times the difference between both is 
semiclassically negligible.

\section{TDWKB approach}

It was recently shown that when the underlying classical 
dynamics of a system is chaotic the evolution of a Gaussian 
wavepacket can be described by the standard time-dependent 
WKB (TDWKB) method \cite{toscano-raul-prl}. 
The reason is that, in semiclassical regimes, after some 
short time the Gaussian wavepacket stretches over a classical 
length $\sim \hbar^0$. Thus, it becomes a primitive WKB 
state: 
\be
\label{initialWKB}
\psi_0(q)= A_0(q)  \exp [ i S_0(q) / \hbar ] \;\; ,
\ee
supported by the Lagrangian manifold defined by 
\be
\label{initial-manifold}
p=dS_0/dq\;\;,
\ee
and where the amplitude $A_0(q)$ and the phase $S_0(q)$ are 
smooth function on the quantum scale.
From then on the wavefunction evolves according to the TDWKB 
recipe:
\be
\label{finalWKB}
\psi_t(q) 
\approx \sum_\nu              A^{(\nu)}_t(q)  
                  \exp [ \, i S^{(\nu)}_t(q) / 
\hbar -i\mu_{\nu}\pi/2]  \;,
\ee
where $\nu$ labels the different branches of the Lagrangian 
manifold obtained by evolving classically the initial manifold 
(\ref{initial-manifold}).

In chaotic systems the applicability of TDWKB to wavepackets is
guaranteed by the exponentially fast stretching of phase space
\cite{toscano-raul-prl}. 
We show here that the same scheme can be applied in the case of
{\em integrable nonlinear} systems, where the stretching is 
linear in time.

For the present analysis we found more convenient to make a
slight modification of the Hamiltonian of previous
sections:
\be
\label{2nd-quantum-Hamiltonian}
\hat{H}=  \gamma \hbar^2 \left (\hat{n} - n_0 \right)^2 \,.
\ee
The corresponding classical Hamiltonian is 
\be
\label{2nd-classical-Hamiltonian}
H(q,p)= \gamma (I - I_0 )^2 \,.
\label{ham2}
\ee
This is equivalent to working with the Hamiltonian 
(\ref{1st-quantum-Hamiltonian}) but in the interaction 
representation, with a free evolution given by the 
harmonic oscillator 
$\hat{H}_0=\hbar\omega' \hat{n}+C$ 
[with $\omega'=2\gamma\hbar(n_0+1/2)$ 
and $C$ an appropriate constant)].
Thus, we eliminate the rotation dynamics of the wavepacket, 
while preserving the nonlinear squeezing. 
This choice simplifies the determination of the initial 
WKB manifold (\ref{initial-manifold}), which now remains 
almost stationary; 
otherwise it would rotate, forcing us to change 
representation from time to time to avoid caustics.
Switching to the interaction representation does not affect 
the semiclassical accuracy of the calculation, because the 
transformation generated by $\hat{H}_0$ is semiclassically 
exact \cite{littlejohn84}. 

The centroid $(q_0,p_0)$ of the initial wavepacket will
be chosen in such a way that $I_0=\hbar(n_0+1/2)=(q_0+p_0)/2$.
The numerical implementation of the TDWKB recipe 
\cite{toscano-raul-prl} requires the determination of the 
initial manifold through Eq.~(\ref{initial-manifold}). 
The initial action $S_0(q)$ is extracted from the phase of the 
exact wavefunction propagated up to some short time $t_i$. 
The only condition this time must satisfy is that the exact
wavefunction must be described to good accuracy by a primitive 
WKB state, {\it i.e.},  Eq.~(\ref{initialWKB}) with 
$A(q)$ and $S(q)$ smooth on the quantum scale.

\begin{figure}[htp]
\hspace{0.0cm}
\includegraphics[angle=0.0, width=8cm]{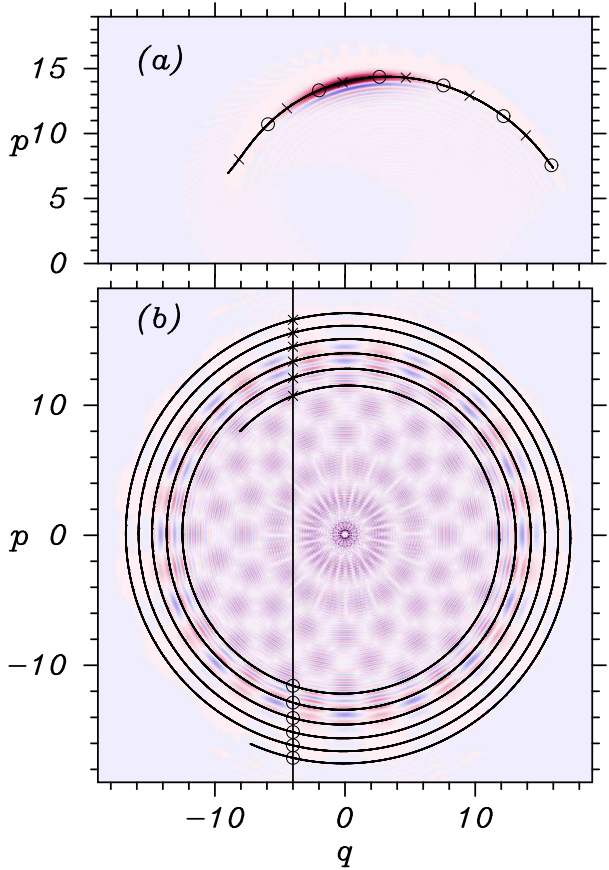}
\caption{{\bf (a)} 
(Color online) Wigner function of the evolved initial 
coherent state, centered at the point $(q_0,p_0)=(0,14)$,
for a time $t=T_2/320$ ($T_2$ the revival time).
The superimposed black line is the initial Lagrangian 
manifold.   
{\bf (b)} 
Wigner function of the same initial coherent state at $t=T_2/16$. 
The spiraling black line is the classical evolved
manifold that supports the WKB state.
Parameters are $m,\gamma,\hbar=1$.
See text for explanation of the meaning of crosses and circles.}
\label{fig5}
\end{figure}

Our choice of the initial manifold is showed in 
Fig.~\ref{fig5}(a).
%
%
%
Using the classical equations of motion we evolve 
this initial manifold up to the desired final time 
[see Fig.~\ref{fig5}  (b)].
In order to calculate the WKB wavefunction $\psi_t(q)$
we have to determine the classical trajectories 
corresponding to each term in the sum in 
Eq.~(\ref{finalWKB}). 
Those are the trajectories that at times $t_i$ have 
initial conditions 
$(q^{(i)}_j,p^{(i)}_j)$ 
on the initial manifold, and at time $t$ reach the final 
manifold at the points
$(q^{(f)}_j,p^{(f)}_j)$, 
where $q^{(f)}_j=q$ for all values of $j$. 
For example, in Fig.~(\ref{fig5}) we show with crosses 
in panel {\bf (a)} the initial conditions of the 
trajectories that end at points 
$(q^{(f)}_j=q,p^{(f)}_j)$ 
with positive momentum [panel {\bf (b)}], 
and with circles the trajectories that end with a 
negative momentum.  
The classical action for these trajectories can be calculated 
analytically for the Hamiltonian in 
Eq.~(\ref{2nd-classical-Hamiltonian}):
\ba
S_t^{(j)}&=&S_0+\int_{t_i}^{t}(p\dot{q}-H)\; dt=
S_0+\frac{1}{2}(p^{(f)}_jq^{(f)}_j-\nonumber \\
&&-p^{(i)}_jq^{(i)}_j)+(\omega(I_j) I_j-H_j)\Delta t\;\;,
\ea 
where 
$\omega(I_j)\equiv2\gamma(I_j-I_0)$, 
$H_j\equiv\gamma(I_j-I_0)^2$ 
and 
$\Delta t\equiv t-t_i$. 
The Maslov index $\mu$ in Eq.(\ref{finalWKB}) equals 
$+/-$ the number of turning points along a 
clockwise/counterclockwise trajectory 
\cite{littlejohn92}.
The amplitudes $A_t^{(j)}$ in Eq.(\ref{finalWKB})
are calculated from the continuity equation 
\be
\label{TDWKBamplitude}
A_t^{(j)}(q^{(f)}_j=q)=
A_0(q^{(i)}_j)\left|\frac{dq_i}{dq_f}\right|^{1/2}\;\;,
\ee
where $A_0(q^{(i)}_j)$ is the amplitude of the primitive 
WKB wavefunction supported by the initial manifold. 
The factor $|dq_i/dq_f|$ is determined by evolving 
numerically a nearby trajectory.

Note that the relatively hard part of this numerical method 
is the calculation of the points 
$(q^{(i)}_j,p^{(i)}_j)$ and 
$(q^{(f)}_j=q,p^{(f)}_j)$ in the initial and the final 
Lagrangian manifolds, respectively. 
This can be done in a systematic way defining a parameter $s$
running along the initial manifold and calculating the points 
$(q^{(f)}_j,p^{(f)}_j)$ as intersections of the trajectories, 
with parameter $s$ and fixed elapsed time $t$, 
with the phase-space vertical line $q=q^{(f)}$, in a way 
resembling a Poincar\'e section map.
 
\begin{figure}[htp]
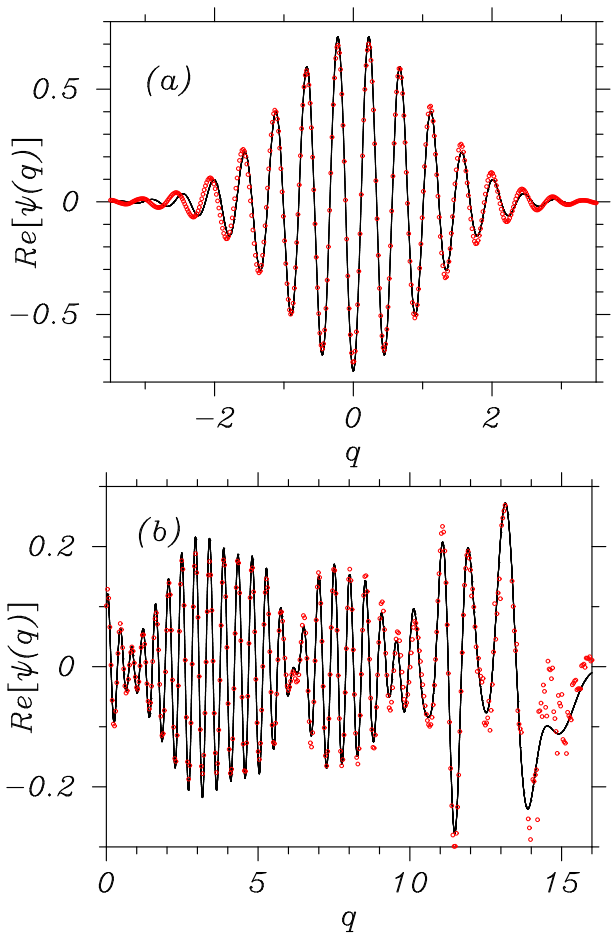

\hspace{0.0cm}
\includegraphics[angle=-90.0, width=8cm]{fig6a.ps}
\includegraphics[angle=-90.0, width=8cm]{fig6b.ps}
\caption{Real part of the evolved wavefunction of an initially 
coherent state centered at the point $(q_0,p_0)=(0,14)$
evolved with the Hamiltonian of 
Eq.~(\ref{2nd-quantum-Hamiltonian}). 
The full line is the exact wavefunction and
the circles correspond to the TDWKB approximation. 
{\bf (a)} For  $t=4T_2$  ($T_2$ is the revival time);
{\bf (b)} For  $T_2/2.38567$. 
Similar agreement is observed for $q<0$ and 
for the imaginary parts.
Parameters are $m,\gamma,\hbar=1$.} 
\label{fig6}
\end{figure}

In Fig.~\ref{fig6} we compare the exact wavefunction of an 
initially coherent state with the TDWKB wavefunction for 
$t=4T_2$  in panel (a) 
and 
for a generic time in panel (b) 
[when the Wigner function is nonlocalized, 
similar to the wavefunction in the rightmost panel of
 Fig.~(\ref{fig1})].
%
%
%
%
In panel (a) we see that the agreement is very good 
for all values of $q$, with small errors for $|q|>1.8$.
For multiples of the revival time, $t=mT_2$, the number of 
classical trajectories needed to build up the TDWKB wavefunction 
grows like $\propto 90\,m$.
Nevertheless, the numerical errors seem to be almost constant.  
Like in the Van Vleck approach of Sec.~III, 
at multiples of the revival time only 
classical trajectories with positive final momenta 
contribute.
 
In  Fig.~\ref{fig6}(b) we show the TDWKB 
approximation in the generic case, 
where trajectories with 
final negative momenta also contribute.
In this example, the exact wavefunction spreads also over 
the regions $-17<q<-11$ and $11<q<17$, where the final manifold  
crosses several times the axis $p=0$.
These points are caustics of the TDWKB approximation, where 
the amplitude in Eq.~(\ref{TDWKBamplitude}) diverges.
Thus, each time the contributing trajectories has a null final 
momentum, the TDWKB approximation breaks down. 
This is clearly seen in Fig~  \ref{fig6} (b).
The proper treatment of the WKB function in this region
requires more sophisticated approximations.
For our present purposes, it is enough to verify that
for $|q|<10$, where there are no caustic points, the 
agreement is excellent.

\section{Concluding remarks}

%

Quantum revival is a subtle phenomenon where interference 
plays a crucial role. 
We studied revivals in the quartic oscillator from 
a semiclassical perspective. 
Among various semiclassical theories existing in the 
literature we chose two of most basic and popular: 
Van Vleck propagation and time-dependent WKB.
%
%
In both cases the results were impressive: quantum dynamics
--in particular, revivals-- can be semiclassically described 
with great accuracy. 
 
In the TDWKB approach, we computed the wavefunction numerically 
for times beyond the first revivals. 
Excluding an initial stage (where standard WKB fails), the 
classical skeleton of the wavefunction appeared to be a 
spiraling manifold.
Thus, we have exhibited another successful test of the TDWKB 
scheme for the propagation of wavepackets 
\cite{toscano-raul-prl} --in this occasion for the long-time 
dynamics
of an integrable system.
However, at present, we cannot assess analytically how 
far the agreement will extend. 
In order to do this one should integrate the present scheme 
with a theory capable of describing the short time dynamics.
The natural tool is {\em complex} TDWKB which uses 
manifolds in the complexified phase space \cite{complexWKB}. 
Then one should prove that, provided the dynamics is 
stretching, the complex manifold describing a coherent
state eventually decays into a real manifold 
[like that shown in Fig.~5(a)].
 

\vspace{1pc}
Like in some previous studies \cite{Mallalieu_Stroud,Barnes}, 
we obtained an expression for 
the autocorrelation function as a sum over
classical trajectories, all the ingredients being given
in closed form.
Furthermore, we showed analytically that the Van Vleck 
correlation function and the exact one are essentially
equal.
The key step was to use the Poisson transformation, which 
reshapes a semiclassical correlation function into a 
quantum-looking one, or vice versa.
Similar analyses may be possible both for the Coulomb 
potential \cite{Mallalieu_Stroud,Barnes} and the Morse 
oscillator \cite{Wang-Heller} 
(the latter is simpler, in principle, because its
spectrum is quadratic in the quantum 
number \cite{child}).

The system under study, the quartic oscillator, is
special, even among integrable systems, in that its 
Hamiltonian is a (quadratic) function of the action 
variable.
Thus, for instance, stationary WKB theory gives the 
exact energy levels for this system.
We have provided analytical and numerical evidence 
showing that the semiclassical {\em time-dependent} 
schemes considered in this paper are also ``exact".
It remains to ascertain if these considerations extend
to more general $H(I)$ Hamiltonians, e.g., of the
polinomial type.

\vspace{1pc}
Concerning the work by Novaes \cite{Novaes}, who studied 
the quartic oscillator by using the semiclassical 
coherent-state representation of the propagator, it is 
likely, in the light of our results, that it should be 
possible to identify the relevant subset of complex 
trajectories which contribute to the autocorrelation 
function at long times.

%
%

%
\section*{Acknowledgments}

%
We thank 
    A. M. Ozorio de Almeida, 
    M. Saraceno,
    M. A. M. de Aguiar,
and
    M. Novaes
for useful comments.
Partial financial support from 
ANPCyT, CNPq, CONICET (PIP-6137), PROSUL, and  UBACyT (X237).
is gratefully acknowledged. 
D. W. is a researcher of CONICET.


\end{document}